\documentclass[twocolumn,aps,showpacs,preprintnumbers,amsmath,amssymb,nofootinbib]{revtex4}

\usepackage[english]{babel}
\usepackage{graphicx}
\usepackage{dcolumn}
\usepackage{bm}
\usepackage{verbatim}

\newcommand{\ket}[1]{\left| {#1} \right\rangle}
\newcommand{\bra}[1]{\left\langle {#1} \right|}

\newcommand{\proj}[2]{\left| {#1} \right\rangle\!\left\langle {#2} \right|}

\newcommand{\biket}[2]{\left| {#1} \right\rangle_{I}\left| {#2} \right\rangle_{IV}}
\newcommand{\biketn}[2]{\left| {#1}_k \right\rangle_{I}^+\left| {#2}_k \right\rangle_{IV}^-}
\newcommand{\pa}{p}
\newcommand{\tr}{\operatorname{Tr}}

\def\slashchar#1{\setbox0=\hbox{$#1$} 
\dimen0=\wd0 
\setbox1=\hbox{/} \dimen1=\wd1 
\ifdim\dimen0>\dimen1 
\rlap{\hbox to \dimen0{\hfil/\hfil}} 
#1 
\else 
\rlap{\hbox to \dimen1{\hfil$#1$\hfil}} 
/ 
\fi}

\begin{document}


\title{Quantum correlations through event horizons: Fermionic vs bosonic entanglement}
\author{Eduardo Mart\'{i}n-Mart\'{i}nez}%
 \email{martin@imaff.cfmac.csic.es}
\author{Juan Le\'on}
\email{leon@imaff.cfmac.csic.es}
 \homepage{http://www.imaff.csic.es/pcc/QUINFOG/}
\affiliation{%
Instituto de F\'{i}sica Fundamental, CSIC\\
Serrano 113-B, 28006 Madrid, Spain.\\
}


\date{\today}

\begin{abstract}
We  disclose the behaviour of quantum and classical correlations among all the different spatial-temporal regions of a space-time with an event horizon, comparing fermionic with bosonic fields. We show the emergence of conservation laws for entanglement and classical correlations, pointing out the crucial role that statistics plays in the information exchange (and more specifically, the entanglement tradeoff) across horizons. The results obtained here could shed new light on the problem of  information behaviour in non-inertial frames and in the presence of horizons, giving a better insight about the black hole information paradox. 
\end{abstract}

\maketitle


\section{Introduction}

Relativistic quantum information, among other topics,  analyses entanglement behaviour in non-inertial settings.  It combines tools from general relativity, quantum field theory and quantum information theory. It is a new and fast-growing field \cite{Alsingtelep,TeraUeda2,ShiYu,Alicefalls,AlsingSchul,SchExpandingspace,Adeschul,KBr,LingHeZ,ManSchullBlack,PanBlackHoles,AlsingMcmhMil,DH,Steeg,Edu2, schacross}. Amongst its hot topics is the analysis of how the Unruh effect \cite{DaviesUnr,Unruh,Takagi,Crispino}  affects the possible entanglement that an accelerated observer would share with an inertial observer. 

For a bipartite entangled system, it is commonplace in relativistic quantum information to call the two observers Alice and Rob. We now consider that while Alice proper frame is inertial, Rob undergoes a constant acceleration $a$.

It was shown \cite{Alicefalls,AlsingSchul,Edu2} that the Unruh effect degrades the entanglement between the two partners affecting all the quantum information tasks that they could perform. Specifically, it was demonstrated that, as Rob accelerates, entanglement is completely degraded for a scalar field and, conversely, some degree of entanglement is preserved for fermionic fields. This behaviour of fermionic fields has been proven to be universal \cite{Edu3}, namely, it is independent of  i) the spin of the fermionic field,  ii) the kind of maximally entangled state from which we start, and iii) the number of participating modes when going beyond the single mode approximation (even an infinite number of them).

When Rob accelerates, the description of his partial state must be done by means of Rindler coordinates \cite{gravitation,Takagi}. As it will be shown below, when doing that the description of the system splits in three different subsystems; Alice's Minkowskian system, a subsystem in region $I$ of Rindler space-time (which we assign to Rob) and another subsystem, called AntiRob, constituted by the modes of the field in region $IV$ of Rindler space time.

It is important to notice that Schwarzschild metrics in the neighbourhood of the event horizon can be approximated by Rindler metrics. Therefore, an observer arbitrarily close to the Schwarzschild event horizon would correspond to an observer arbitrarily close to the Rindler horizon. Being arbitrarily close to the Rindler horizon is achieved when the acceleration parameter $a$ goes to infinity. This would mean that the  limit $a\rightarrow\infty$ of our analysis would correspond to a scenario in which Rob is resisting arbitrarily close to the event horizon of an Schwarzschild black hole, while Alice is free-falling into it.  

Any accelerated observer is constrained to either region $I$ or $IV$ of Rindler space-time. If we select region $I$ coordinates to account for the accelerated observer Rob, he would remain causally disconnected from region $IV$, and therefore, Rob would be unable to communicate with the hypothetical observer AntiRob in region $IV$.

That means that to describe the point of view of Rob we need to remove the part of the system on the other side of the event horizon. This is done by tracing over region $IV$ of Rindler space-time, in other words, erasing the AntiRob information from the system. This partial tracing is the final responsible for the entanglement degradation due to Unruh effect. 

To gain a deeper understanding of those degradation mechanisms it is useful to study how the entanglement is lost as one traces over regions of the Rindler space-time. Although the system is obviously bipartite (Alice and Rob), shifting to Rindler coordinates for Rob the mathematical description of the system \cite{Alicefalls,AlsingSchul,Edu2} admits a straightforward tripartition: Minkowskian modes (Alice), Rindler region $I$ modes (Rob), and Rindler region $IV$  modes (AntiRob). 

In this paper, instead of considering only the Alice-Rob bipartition, we deal with all the different bipartitions of the system to study the correlations tradeoff among them. These three bipartitions are\begin{enumerate}
\item Alice-Rob $(AR)$
\item Alice-AntiRob $(A\bar R)$
\item Rob-AntiRob $(R \bar R)$
\end{enumerate}

Bipartition number 1 is the most commonly considered in the literature. It represents the system formed by an inertial observer and the modes of the field which an accelerated observer is able to access.

The second bipartition represents the subsystem formed by the inertial observer Alice and the modes of the field which Rob is not able to access due to the presence of an horizon as he accelerates. The physical meaning of this bipartition can be more clearly understood in the limit $a\rightarrow\infty$, in which Rob  is equivalent to an observer standing outside but arbitrarily close to a Schwarzschild black hole event horizon. Then, AntiRob subsystem represents the field modes inside the event horizon.

The third bipartition lacks physical meaning in terms of information theory because communication between Rob and AntiRob is not allowed. Anyway, studying this bipartition is still useful to account for the correlations which are created between the spatial-temporal regions separated by an event horizon and, therefore, its study is necessary and complementary to the previous ones in order to give a complete description of the information behaviour across an event horizon.

In \cite{AlsingSchul} the existence of these three possible bipartitions was considered only for spinless fermion fields. In this work we will go far beyond previous analysis and we will compare the correlations tradeoff among different bipartitions for bosonic and fermionic fields, showing the leading role of statistics in the behaviour of information on the proximity of event horizons. 

Dimension of the Hilbert space for each mode has been often blamed as responsible for the difference between fermionic and bosonic entanglement behaviour in the presence of horizons. Here we will disclose in which cases changing the dimension affects the correlations behaviour, showing that, for the physical systems, it  is largely irrelevant .

The work presented here will show that the role played by statistics in the comportment of information in the proximity of an event horizon is so important that could even give a hint about the relationship between statistics and the black hole information paradox.

This paper is organised as follows, in section \ref{sec2} we introduce some basic notions about an accelerated observer reference frame and present the Bogoliubov transformations which relate the Minkowskian modes of the fields with its analogous in Rindler coordinates. In section \ref{sec3} we introduce some notation and present the vacuum and one particle states for a scalar and a Dirac field as seen from an accelerated observer point of view. We also write down the qubit states which we are going to analyse when one of the partners accelerates. In section \ref{sec4} we compute the quantum and classical correlations (in terms of mutual information and measures of quantum entanglement) for all the possible bipartitions of the system for a Dirac field, showing the emergence of an entanglement conservation law for the systems $AR$ and $A\bar R$, as well as the conservation of classical correlations. In section \ref{sec5} we repeat the same exercises for the case of scalar fields, finding striking differences  which point out the enormous impact of statistics on correlations behaviour. In both, section \ref{sec4} and \ref{sec5} we also analyse the correlations across the horizon, i.e., the $R\bar R$ system. Then we present our results and conclusions in section \ref{conclusions}.

\section{Scalar and Dirac fields from constantly accelerated frames}\label{sec2}

An uniformly accelerated observer viewpoint is described by means of the Rindler coordinates \cite{gravitation}. In order to cover the whole Minkowski space-time, two different sets of coordinates are necessary. These sets of coordinates define two causally disconnected regions in Rindler space-time. If we consider that the uniform acceleration $a$ lies on the $z$ axis, the new Rindler coordinates $(t,x,y,z)$ as a function of Minkowski coordinates $(\tilde t,\tilde x,\tilde y,\tilde z)$ are
\begin{equation}\label{Rindlcoordreg1}
a\tilde t=e^{az}\sinh(at),\; a\tilde z=e^{az}\cosh(at),\; \tilde x= x,\; \tilde y= y
\end{equation}
for region I, and
\begin{equation}\label{Rindlcoordreg2}
a\tilde t=-e^{az}\sinh(at),\; a\tilde z=-e^{az}\cosh(at),\; \tilde x= x,\; \tilde y= y
\end{equation}
for region IV.
\begin{figure}\label{fig1}
\includegraphics[width=.45\textwidth]{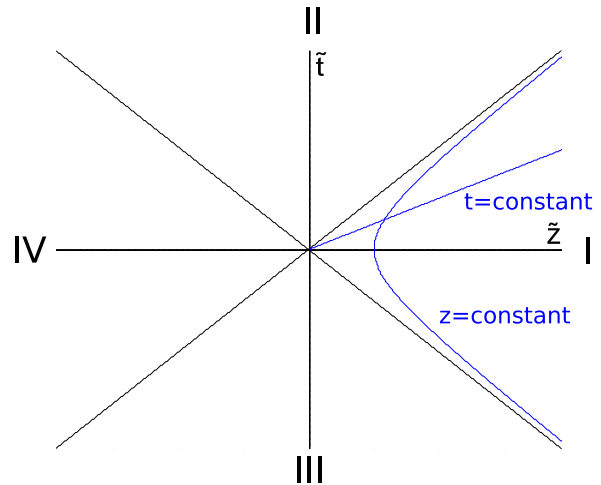}
\caption{(Colour online) Rindler space-time diagram: lines of constant position $z=\text{const.}$ are hyperbolae and all the curves of constant proper time $t$ for the accelerated observer are straight lines that come from the origin. An uniformly accelerated observer Rob travels along a hyperbola constrained to region I}
\end{figure}
As we can see from fig. 1, although we have covered the whole Minkowski space-time with these sets of coordinates, there are two more regions labeled II and III. To map them we would need to trade $\cosh$ by $\sinh$ in equations \eqref{Rindlcoordreg1},\eqref{Rindlcoordreg2}. In these regions, $t$ is a spacelike coordinate and $z$ is a timelike coordinate. However,  considering such regions is not required to describe fields from an accelerated observer perspective \cite{Birrell,gravitation,Alicefalls,AlsingSchul,Edu2}.

The Rindler coordinates $z,t$ go from $-\infty$ to $\infty$ independently in regions I and IV. It means that each region admits a separate quantization procedure with their corresponding positive and negative energy solutions of Klein-Gordon or Dirac equations\footnote{Throughout this work we will consider that the spin of each mode lies on the acceleration direction and, hence, spin will not undergo Thomas precession due to instant Wigner rotations \cite{AlsingSchul,Jauregui}.} $\{\psi^{I+}_{k,s},\psi^{I-}_{k,s}\}$ and $\{\psi^{IV+}_{k,s},\psi^{IV-}_{k,s}\}$.

Particles and antiparticles will be classified with respect to the future-directed timelike Killing vector in each region. In region I the future-directed Killing vector is
\begin{equation}\label{KillingI}
\partial_t^I=\frac{\partial \tilde t}{\partial t}\partial_{\tilde t}+\frac{\partial\tilde z}{\partial t}\partial_{\tilde z}=a(\tilde z\partial_{\tilde t}+\tilde t\partial_{\tilde z}),
\end{equation}
whereas in region IV the future-directed Killing vector is $\partial_t^{IV}=-\partial_t^{I}$.

This means that solutions in region I, having time dependence $\psi_k^{I+}\sim e^{-ik_0t}$ with $k_0>0$, represent positive energy solutions, whereas solutions in region IV, having time dependence $\psi_k^{I+}\sim e^{-ik_0t}$ with $k_0>0$, are actually negative energy solutions since $\partial^{IV}_t$ points to the opposite direction of $\partial_{\tilde t} $. As I and IV are causally disconnected $\psi^{IV\pm}_{k,s}$ and $\psi^{I\pm}_{k,s}$ only have support in their own regions, vanishing outside them.

Let us denote $(a_{I,k}^{\phantom{\dagger}},a^{\dagger}_{I,k})$ for the scalar field  and $(b^{\phantom{\dagger}}_{I,k,s},b^{\dagger}_{I,k,s})$ for the Dirac field as the particle annihilation and creation operators in region I, and $(c^{\phantom{\dagger}}_{I,k,s},c^{\dagger}_{I,k,s})$ the corresponding antiparticle Dirac field operators. Analogously we define $(a^{\phantom{\dagger}}_{IV,k},a^{\dagger}_{IV,k},b^{\phantom{\dagger}}_{IV,k,s},b^{\dagger}_{IV,k,s}, c^{\phantom{\dagger}}_{IV,k,s},c_{IV,k,s}^\dagger)$ as the particle/antiparticle operators in region IV.

The bosonic operators satisfy the commutation relations $[a^{\phantom{\dagger}}_{\text{R},k},a^\dagger_{\text{R}',k'}]=\delta_{\text{R}\text{R}'}\delta_{kk'}$ and the fermionic operators satisfy the anticommutation relations $\{c^{\phantom{\dagger}}_{\text{R},k,s},c^\dagger_{\text{R}',k',s'}\}=\delta_{\text{R}\text{R}'}\delta_{kk'}\delta_{ss'}$. The subscript R notates the Rindler region of the operator $\text{R}=\{I,IV\}$.  All other commutators and anticommutators are zero. This includes the anticommutators between operators in different regions of the Rindler space-time.

We can relate Minkowski and Rindler creation and annihilation operators by taking appropriate inner products and computing the so-called Bogoliubov coefficients \cite{Takagi,Jauregui,Birrell,AlsingSchul}.

For a scalar field, the Bogoliubov relationships for the annihilation operator of modes with positive frequency are
\begin{eqnarray}\label{bogoboson}
 a_{M,k}&=&\cosh r_s\, a_{I,k} - \sinh r_s\, a^\dagger_{IV,-k}
\end{eqnarray}
where
\begin{equation}\label{defr1}
\tanh r_s=e^{-\pi \frac{k_0c}{a}}
\end{equation}

For a Dirac field, the Bogoliubov relationships take the form
\begin{eqnarray}\label{bogodirac}
\nonumber b_{M,k,s}&=&\cos{r_d}\,b_{I,k,s}-\sin r_d\,c^\dagger_{IV,-k,-s}\\*
c^\dagger_{M,k,s}&=&\cos{r_d}\,c^\dagger_{IV,k,s}+\sin r_d\,b_{I,-k,-s}
\end{eqnarray}
where
\begin{equation}\label{defr2}
\tan r_s=e^{-\pi \frac{k_0c}{a}}
\end{equation}

\section{Vacuum and one particle states}\label{sec3}

The Minkowski vacuum state for the scalar field is defined by the tensor product of each frequency mode vacuum
\begin{equation}\label{vacuas}\ket0=\bigotimes_{k}\ket{0_k}\end{equation}
such that it is annihilated by $a_{k}$ for all values of $k$.

The Minkowski vacuum state for the Dirac field is defined by the tensor product of each frequency mode vacuum
\begin{equation}\label{vacuad}\ket0=\bigotimes_{\substack{k,k'\\s,s'}}\ket{0_{k,s}}^+\ket{0_{k',s'}}^-\end{equation}
such that it is annihilated by $b_{k,s}$ and $c_{k,s}$ for all values of $k, s$. The $\pm$ label indicates particle/antiparticle  mode.

For the sake of this work we are going to constrain ourselves to the single mode approximation (SMA) \cite{Alsingtelep,AlsingMcmhMil}. In any case in \cite{Edu3} we showed, as an universality principle, that going beyond this approximation does not modify the way in which Unruh decoherence affects entanglement of spinless fermionic and Dirac fields. Specifically, it was shown that Unruh decoherence is independent of the number of modes of the field considered in the analysis, being  statistics the ruler of this process. Hence, for our purposes, carrying out this approximation or not will not be relevant.

As it is shown in \cite{Alicefalls}, the vacuum state for a $k$-momentum mode of a scalar field seen from the perspective of an accelerated observer is
\begin{equation}\label{scavac}
\ket{0_k}_M=\frac{1}{\cosh r_s}\sum_{n=0}^\infty \tanh^n r_s \ket{n_k}_I\ket{n_{-k}}_{IV}
\end{equation}
and as it is shown in \cite{Edu2}, the vacuum state for a Dirac field seen from the accelerated frame is
\begin{eqnarray}\label{diravac}
\nonumber \ket{0}_M&=&\cos^2 r_d\,\biketn{0}{0}+\sin r_d\,\cos r_d\left(\biketn{\uparrow}{\downarrow}\right.\\*
&&\left.+\biketn{\downarrow}{\uparrow}\right)+\sin^2 r_d\,\biketn{\pa}{\pa}
\end{eqnarray}
where $\ket{\pa_k}^\pm$ represents the pair of particles/antiparticles for frequency $k$ as defined below. In these expressions we use the same notation as in references \cite{Alicefalls,Edu2}.

Notice that for fermions there is a constraint due to Pauli exclusion principle
\begin{equation}b^\dagger_{k,s}b^\dagger_{k,s'}\ket0=\ket{ss'_k}\delta_{s,{-s'}}\end{equation}
If $s=s'$ the two particle state is not allowed. Therefore the allowed Minkowski states for each mode of particle/antiparticle are
\begin{equation}\{\ket{0_k}^\pm,\ket{\uparrow_k}^\pm,\ket{\downarrow_k}^\pm,\ket{\pa_k}^\pm\}.\end{equation}

From now on we will drop the sign $\pm$ as, in this work, a mode in region $I$ will always be a particle mode and a mode in region $IV$ will always represent an antiparticle mode. To simplify notation we will also drop the $k$ label as we are working under SMA.

We will use the following definitions for a pair of fermions
\begin{eqnarray}\label{notation2}
\nonumber \ket{\pa}_I&=&b^\dagger_{I\uparrow}b^\dagger_{I\downarrow}\ket{0}_I=-b^\dagger_{I\downarrow}b^\dagger_{I\uparrow}\ket{0}_I\\*
\ket{\pa}_{IV}&=&c^\dagger_{IV\uparrow}c^\dagger_{IV\downarrow}\ket{0}_{IV}=-c^\dagger_{IV\downarrow}c^\dagger_{IV\uparrow}\ket{0}_{IV}
\end{eqnarray}
and, being consistent with the different Rindler regions operators anticommutation relations,
\begin{equation}\label{notation3}
\nonumber \ket{s}_I\ket{s'}_{IV}=b^\dagger_{Is}c^\dagger_{IVs'}\ket{0}_{I}\ket{0}_{IV}=-c^\dagger_{IVs'}b^\dagger_{Is}\ket{0}_{I}\ket{0}_{IV}
\end{equation}
\begin{equation}\label{notation4}
c^\dagger_{IVs'}\biket{s}{0}=-\biket{s}{s'}.
\end{equation}

For this work we will also need the Minkowskian one particle state in Rindler coordinates. 
This state would be 
\begin{equation}
\ket{1}_M=\frac{1}{\cosh^2 r_s}\sum_{n=0}^\infty \tanh^n r_s \,\sqrt{n+1}\ket{n+1}_I\ket{n}_{IV}
\end{equation} 	
for the scalar field \cite{Alicefalls} and 
\begin{eqnarray}\label{onepart2}
\nonumber\ket\uparrow_M&=&\cos r_d \biket{\uparrow}{0}+\sin r_d\biket{\pa}{\uparrow}\\*
\ket\downarrow_M&=&\cos r_d \biket{\downarrow}{0}-\sin r_d\biket{\pa}{\downarrow}
\end{eqnarray}
for the Dirac field \cite{Edu2}.

Now we need to consider the following maximally entangled states in Minkowsky coordinates
\begin{eqnarray}
\label{entangledsca}\ket{\Psi_s}&=&\frac{1}{\sqrt{2}}\left(\ket{0}_M\ket{0}_M+\ket{1}_M\ket{1}_M\right)\\*
\label{entangleddir}\ket{\Psi_d}&=&\frac{1}{\sqrt{2}}\left(\ket{0}_M\ket{0}_M+\ket{\uparrow}_M\ket{\downarrow}_M\right)
\end{eqnarray}
These two maximally entangled states are analogous, both are qubit states and superpositions of the bipartitie vacuum and the bipartite one particle state. The difference is that in \eqref{entangleddir} we have a Dirac field state and hence, the one particle states have spin.

For $\ket{\Psi_d}$ we have selected one amongst the possible values for the spin of the terms with one particle for Alice and Rob, but it can be shown that the election of a specific value for these spins is not relevant when considering the behaviour of corrleations. Then, the results presented here are independent of the particular choice of a spin state for the superposition \eqref{entangleddir}.

\section{Correlations for the Dirac field}\label{sec4}

The density matrix for whole tripartite state, which includes modes on both sides of the Rindler horizon along with Minkowskian modes, is built from \eqref{entangleddir}
\begin{equation}\label{tripadir}
\rho^{AR\bar R}_d=\proj{\Psi_d}{\Psi_d}
\end{equation}

The three different  bipartitions for the Dirac field case are obtained by partial tracing over the part which we want to eliminate, which is to say.
\begin{eqnarray}
\label{AR2}\rho^{AR}_d&\!\!=\!&\tr_{IV}\rho^{AR\bar R}_d=\!\!\!\!\sum_{s\in\{0,\uparrow,\downarrow,\pa\}} \bra{s}_{IV}\rho_d^{AR\bar R}\ket{s}_{IV}\\*
\label{AAR2}\rho^{A\bar R}_d&\!\!=\!&\tr_{I}\rho^{AR\bar R}_d=\!\!\!\!\sum_{s\in\{0,\uparrow,\downarrow,\pa\}} \bra{s}_{I}\rho_d^{AR\bar R}\ket{s}_{I}\\*
\label{RAR2}\rho^{R\bar R}_d&\!\!=\!&\tr_{M}\rho^{AR\bar R}_d=\!\!\!\!\sum_{s\in\{0,\uparrow,\downarrow,\pa\}} \bra{s}_{M}\rho_d^{AR\bar R}\ket{s}_{M}
\end{eqnarray}
and the density matrix for each individual subsystem is obtained by tracing over the other subsystems,
 \begin{eqnarray}
\label{A2}\rho^{A}_d&=&\tr_{I}\rho^{AR}_d=\tr_{IV}\rho^{A\bar R}_d\\*
\label{R2}\rho^{R}_d&=&\tr_{IV}\rho^{R\bar R}_d=\tr_{M}\rho^{AR}_d\\*
\label{aR2}\rho^{\bar R}_d&=&\tr_{I}\rho^{R\bar R}_d=\tr_{M}\rho^{A \bar R}_d
\end{eqnarray}

The different bipartitions are characterized by the following density matrices
\begin{eqnarray}\label{rhoars2}
\nonumber \rho^{AR}_d&=&\frac12\Big[\cos^4 r_d \proj{00}{00}+\sin^2 r_d\,\cos^2 r_d\Big(\proj{0\uparrow}{0\uparrow}\\*
&&\!\!\!\nonumber+\proj{0\downarrow}{0\downarrow}\Big)+\sin^4 r_d\proj{0\pa}{0\pa}+\cos^3r_d\Big(\ket{00}\nonumber\\*
&&\!\!\! \times\bra{\uparrow\downarrow}  +\proj{\uparrow\downarrow}{00}\Big)-\sin^2 r_d\cos r_d\Big(\proj{0\uparrow}{\uparrow\pa}\nonumber\\*
&& +\proj{\uparrow\pa}{0\uparrow}\Big)+\cos^2 r_d \proj{\uparrow\downarrow}{\uparrow\downarrow}+\sin^2 r_d\nonumber\\*
&&\times\proj{\uparrow\pa}{\uparrow\pa}\Big]
\end{eqnarray}
\begin{eqnarray}\label{rhoa-rs2}
\nonumber \rho^{A\bar R}_d&=&\frac12\Big[\cos^4 r_d \proj{00}{00}+\sin^2 r_d\,\cos^2 r_d\Big(\proj{0\downarrow}{0\downarrow}\\*
&&\!\!\!\nonumber+\proj{0\uparrow}{0\uparrow}\Big)+\sin^4 r_d\proj{0\pa}{0\pa}-\sin^3r_d \nonumber\\*
&&\!\!\! \times \Big(\proj{0\pa}{\uparrow\downarrow}+\proj{\uparrow\downarrow}{0\pa}\Big)+\sin r_d\cos^2 r_d\Big(\ket{0\uparrow}\nonumber\\*
&&\times\bra{\uparrow0}+\proj{\uparrow0}{0\uparrow}\Big)+\cos^2 r_d \proj{\uparrow0}{\uparrow0}\nonumber\\*
&&+\sin^2 r_d\proj{\uparrow\downarrow}{\uparrow\downarrow}\Big]
\end{eqnarray}
\begin{eqnarray}\label{rhor-rs2}
\nonumber \rho^{R\bar R}_d\!\!\!&=&\!\!\!\frac{1}{2}\Big[\cos^4 r_d \proj{00}{00}+\sin r_d\, \cos^3 r_d\Big(\proj{00}{\uparrow\downarrow}+\ket{00}\\*
&&\!\!\!\!\!\nonumber\times \bra{\downarrow\uparrow}+\proj{\uparrow\downarrow}{00}+\proj{\downarrow\uparrow}{00}\Big)+\sin^2 r_d\cos^2 r_d\Big(\ket{00}\\*
&&\!\!\!\!\!\nonumber\times\bra{\pa\pa}\!+\!\proj{\uparrow\downarrow}{\uparrow\downarrow}\!+\!\proj{\uparrow\downarrow}{\downarrow\uparrow}\!+\!\proj{\downarrow\uparrow}{\uparrow\downarrow}\!+\!\proj{\downarrow\uparrow}{\downarrow\uparrow}\\*
&&\!\!\!\!\!\nonumber +\proj{\pa\pa}{00}\Big)+ \sin^3 r_d\,\cos r_d\Big(\proj{\uparrow\downarrow}{\pa\pa}\\*
&&\!\!\!\!\!\nonumber+\proj{\pa\pa}{\uparrow\downarrow}+\ket{\downarrow\uparrow}\bra{\pa\pa}\!+\!\proj{\pa\pa}{\downarrow\uparrow}\Big)\\*
&&\!\!\!\!\!\nonumber+\cos^2 r_d\proj{\downarrow 0}{\downarrow 0}+\sin^2 r_d\proj{\pa\downarrow}{\pa\downarrow}\\*
&&\!\!\!\!\!\nonumber-\cos r_d\,\sin r_d\Big(\proj{\downarrow0}{\pa\downarrow}+\ket{\pa\downarrow}\bra{\downarrow0}\Big)\\*
&&\!\!\!\!\!+\sin^4r_d\proj{\pa\pa}{\pa\pa}\Big]
\end{eqnarray}
where the bases are
\begin{eqnarray}\label{barbolbasis2}
 \ket{nm}&=&\ket{n^A}_{M}\ket{m^R}_{I}\\*
\ket{nm}&=&\ket{n^A}_{M}|m^{\bar R}\rangle_{IV}\\*
\ket{nm}&=&\ket{n^R}_{I}|m^{\bar R}\rangle_{IV}
\end{eqnarray}
respectively for \eqref{rhoars2}, \eqref{rhoa-rs2} and \eqref{rhor-rs2}.

On the other hand, the density matrices for the individual subsystems \eqref{A2}, \eqref{R2},\eqref{aR2} are
\begin{eqnarray}\label{robfpartialstate}
\nonumber \rho^{R}_d&=&\frac12\Big[\sin^2r_d(1+\sin^2 r_d)\proj{\pa}{\pa}+\sin^2 r_d\cos^2r_d\proj{\uparrow}{\uparrow}\\* 
&&+\cos^2 r_d(1+\sin^2 r_d)\proj{\downarrow}{\downarrow}+\cos^4r_d\proj{0}{0}\Big]
\end{eqnarray}
\begin{eqnarray}\label{arobfpartialstate}
\nonumber \rho^{\bar R}_d&=&\frac12\Big[\cos^2r_d(1+\cos^2r_d)\proj{0}{0}+\sin^2 r_d\cos^2r_d\proj{\uparrow}{\uparrow}\\* 
&&+\sin^2 r_d(1+\cos^2 r_d)\proj{\downarrow}{\downarrow}+\sin^4r_d\proj{\pa}{\pa}\Big]
\end{eqnarray}
\begin{equation}\label{alicefpartialstate}
 \rho^{A}_d=\frac12\left(\proj{0}{0}+\proj{\uparrow}{\uparrow}\right)
\end{equation}

\subsection{Mutual Information: creation, exchange and conservation}

Mutual information accounts for correlations (both quantum and classical) between two different parts of a system. It is defined as
\begin{equation}\label{mutualdef}
I_{AB}=S_A+S_B-S_{AB}
\end{equation}
where $S_A$, $S_B$ and $S_{AB}$ are respectively the Von Neuman entropies for the individual subsystems $A$ and $B$ and for the joint system $AB$.

To compute the mutual information  for each bipartition we will need the eigenvalues of the corresponding density matrices. We shall go through all the process step by step in the lines below.

\subsubsection{Bipartition Alice-Rob}

The eigenvalues of the matrix for the system Alice-Rob \eqref{rhoars2} are
\begin{eqnarray}\label{eigAR1}
\nonumber \lambda_1&=&\lambda_2=0\\*
\nonumber \lambda_3&=&\frac12\sin^2r_d\cos^2r_d\\*
\nonumber \lambda_4&=&\frac12\sin^4r_d\\*
\nonumber \lambda_5&=&\frac12\cos^2r_d\left(1+\cos^2r_d\right)\\*
\lambda_6&=&\frac12\sin^2r_d\left(1+\cos^2r_d\right)
\end{eqnarray}

\subsubsection{Bipartition Alice-AntiRob}

The eigenvalues of the matrix for the system Alice-AntiRob \eqref{rhoa-rs2} are
\begin{eqnarray}\label{eigAaR1}
\nonumber \lambda_1&=&\lambda_2=0\\*
\nonumber \lambda_3&=&\frac12\sin^2r_d\cos^2 r_d\\*
\nonumber \lambda_4&=&\frac12\cos^4 r_d\\*
\nonumber \lambda_5&=&\frac12\sin^2r_d \left(1+\sin^2 r_d\right)\\*
 \lambda_6&=&\frac12\cos^2 r_d\left(1+\sin^2 r_d\right)
\end{eqnarray}

\subsubsection{Bipartition Rob-AntiRob}

All the eigenvalues of the matrix for the system Rob-AntiRob \eqref{rhor-rs2} are zero excepting two of them
\begin{equation}\label{eigRaR1}
\lambda_1=\lambda_2=\frac12;\qquad \lambda_{i>2}=0
\end{equation}

\subsubsection{Von Neumann entropies for each subsystem and mutual information}

To compute the Von Neumann entropies we need the eigenvalues of every bipartition and the individual density matrices. The eigenvalues of $\rho^{AR}_d$, $\rho^{A\bar R}_d$, $\rho^{R\bar R}_d$ are respectively \eqref{eigAR1}, \eqref{eigAaR1} and \eqref{eigRaR1}.

The eigevalues of the individual systems density matrices can be directly read from \eqref{robfpartialstate}, \eqref{arobfpartialstate} and \eqref{alicefpartialstate} since $\rho^R_s$, $\rho^{\bar R}_s$ and $\rho^A_s$ have diagonal forms in the given basis.

The Von Neumann entropy for a partition $B$ of the system is
\begin{equation}\label{Vonneu}
S_B=-\tr(\rho\log_2 \rho)=-\sum \lambda_i\log_2\lambda_i
\end{equation}

At this point, computing the entropies is quite straightforward. The Von Neumann entropies for all the partial systems are
\begin{eqnarray}
\nonumber S_R&=&1-\sin^2r_d\log_2(\sin^2r_d)-\frac32\cos^2r_d\log_2(\cos^2r_d)\\*
&&\nonumber-\frac{1+\sin^2r_d}{2}\log_2(1+\sin^2r_d)\\*
\nonumber S_{\bar R}&=&1-\cos^2r_d\log_2(\cos^2r_d)-\frac32\sin^2r_d\log_2(\sin^2r_d)\\*
\nonumber&&-\frac{1+\cos^2r_d}{2}\log_2(1+\cos^2r_d)\\*
 S_{AR}&=&S_{\bar R}; \qquad S_{A\bar R}=S_{R}; \qquad S_{R\bar R}=S_{A}=1
\end{eqnarray}

And then, the mutual information for all the possible bipartitions of the system will be
\begin{eqnarray}
\nonumber I_{AR}&=&S_A+S_R-S_{A R}=1+S_R -S_{\bar R}\\*
\nonumber I_{A\bar R}&=&S_A+S_{\bar R}-S_{A\bar R}=1+ S_{\bar R}-S_R\\*
\nonumber I_{R\bar R}&=& S_R+S_{\bar R}-S_{R\bar R}=S_R+S_{\bar R}-1
\end{eqnarray}  
At first glance we can see a conservation law of the mutual information for the system Alice-Rob and Alice-AntiRob
\begin{equation}\label{conservation1}
 I_{AR} + I_{A\bar R}=2
\end{equation}
which suggests a correlations transfer from the system Alice-Rob to Alice-AntiRob as the acceleration increases.

Fig. \ref{mutuferm} shows the behaviour of the mutual information for the three bipartitions. It also shows how the correlations across the horizon (Rob and AntiRob) increase, up to certain finite limit, as Rob accelerates. 

If we recall the results on spinless fermion fields \cite{AlsingSchul} we see that the results for Alice-Rob and Alice-AntiRob are exactly the same as those obtained in \cite{AlsingSchul}, being the conservation law obtained here also valid for that spinless fermion case. This result was expected according to the universality argument adduced  in \cite{Edu3} as the explanation for the Unruh decoherence for fermion fields of arbitrary spin.

However, something different occurs with the system Rob-AntiRob. The creation of correlations between modes on both sides of the horizon is greater in the Dirac field case. This is related with the dimension of the Hilbert space. As we will see in more detail later, the dimension of the Hilbert space plays a determinant role only in the comportment of the correlations Rob-AntiRob. 

\subsection{Entanglement conservation and behaviour across the horizon}

We will use the negativity to account for the distillable entanglement of the different bipartitions of the system. Negativity is an entangle monotone which is only sensitive to distillable entanglement.

Negativity is defined as the sum of the negative eigenvalues of the partial transpose density matrix for the system, which is defined as the transpose of only one of the subsystem q-dits in the bipartite density matrix. If $\sigma_i$ are the eigenvalues of $\rho^{pT}_{AB}$ then
\begin{equation}\label{negativitydef}
\mathcal{N}_{AB}=\frac12\sum_{ i}(|\sigma_i|-\sigma_i)=-\sum_{\sigma_i<0}\sigma_i
\end{equation}

Therefore, to compute it, we will need the partial transpose of the bipartite density matrices \eqref{rhoars2}, \eqref{rhoa-rs2} and \eqref{rhor-rs2}, which we will notate as $\eta^{AR}_d$, $\eta^{A\bar R}_d$ and $\eta^{R \bar R}_d$  respectively.

\begin{eqnarray}\label{etaARd}
\nonumber\eta^{AR}_d&=&\frac12\Big[\cos^4r_d\proj{00}{00}+\sin^2r_d\cos^2r_d\left(\proj{0\uparrow}{0\uparrow}\right.\\*
\nonumber&&\left.+\proj{0\downarrow}{0\downarrow}\right)+\sin^4r_d\proj{0\pa}{0\pa}+\cos^3r_d\\*
\nonumber&&\times\left(\proj{0\downarrow}{\uparrow0}+\proj{\uparrow0}{0\downarrow}\right)-\sin^2r_d\,\cos r_d\\*
\nonumber&&\times\left(\proj{0\pa}{\uparrow\uparrow}+\proj{\uparrow\uparrow}{0\pa}\right)+\cos^2r_d\proj{\uparrow\downarrow}{\uparrow\downarrow}\\*
&&+\sin^2r_d\proj{\uparrow\pa}{\uparrow\pa}\Big]
\end{eqnarray}
\begin{eqnarray}\label{etaAaRd}
\nonumber \eta^{A\bar R}_d&=&\frac12\Big[\cos^4 r_d \proj{00}{00}+\sin^2 r_d\,\cos^2 r_d\Big(\proj{0\downarrow}{0\downarrow}\\*
&&\!\!\!\nonumber+\proj{0\uparrow}{0\uparrow}\Big)+\sin^4 r_d\proj{0\pa}{0\pa}-\sin^3r_d \nonumber\\*
&&\!\!\! \times \Big(\proj{0\downarrow}{\uparrow\pa}+\proj{\uparrow\pa}{0\downarrow}\Big)+\sin r_d\cos^2 r_d\nonumber\\*
&&\times\Big(\ket{00}\bra{\uparrow\uparrow}+\proj{\uparrow\uparrow}{00}\Big)+\cos^2 r_d \proj{\uparrow0}{\uparrow0}\nonumber\\*
&&+\sin^2 r_d\proj{\uparrow\downarrow}{\uparrow\downarrow}\Big]
\end{eqnarray}
\begin{eqnarray}\label{etaRaRd}
\nonumber \eta^{R\bar R}_d\!\!\!&=&\!\!\!\frac{1}{2}\Big[\cos^4 r_d \proj{00}{00}+\sin r_d\, \cos^3 r_d\Big(\proj{0\downarrow}{\uparrow 0}\\*
&&\!\!\!\!\!\nonumber+\ket{0\uparrow} \bra{\downarrow 0}+\proj{\uparrow 0}{0 \downarrow}+\proj{\downarrow 0}{0\uparrow}\Big)+\sin^2 r_d\\*
&&\!\!\!\!\!\nonumber\times \cos^2 r_d\Big(\ket{0\pa}\bra{\pa0}\!+\!\proj{\uparrow\downarrow}{\uparrow\downarrow}\!+\!\proj{\uparrow\uparrow}{\downarrow\downarrow}\\*
&&\!\!\!\!\!\nonumber +\proj{\downarrow\downarrow}{\uparrow\uparrow}+\proj{\downarrow\uparrow}{\downarrow\uparrow}+\proj{\pa0}{0\pa}\Big)+ \sin^3 r_d\\*
&&\!\!\!\!\!\nonumber\times\cos r_d\Big(\proj{\uparrow\pa}{\pa\downarrow}+\proj{\pa\downarrow}{\uparrow\pa}+\ket{\downarrow\pa}\bra{\pa\uparrow}\\*
&&\!\!\!\!\!\nonumber+\proj{\pa\uparrow}{\downarrow\pa}\Big)+\cos^2 r_d\proj{\downarrow 0}{\downarrow 0}+\sin^2 r_d \ket{\pa\downarrow}\\*
&&\!\!\!\!\!\nonumber\times \bra{\pa\downarrow}-\cos r_d\,\sin r_d\Big(\proj{\downarrow\downarrow}{\pa0}+\ket{\pa0}\bra{\downarrow\downarrow}\Big)\\*
&&\!\!\!\!\!+\sin^4r_d\proj{\pa\pa}{\pa\pa}\Big]
\end{eqnarray}

In the following subsections we shall compute the negativity for each bipartition of the system.
 
\subsubsection{Bipartition Alice-Rob}

The eigenvalues of the partial transpose density matrix  for the bipartition Alice-Rob \eqref{etaARd} turn out to be
\begin{eqnarray}\label{eigetaARf}
\nonumber\lambda_1&=&\frac12\cos^4r_d; \qquad \lambda_2=\frac12\cos^2r\sin^2r_d\\*
\nonumber\lambda_3&=&\frac12 \sin^2 r_d; \qquad \lambda_4=\frac12\cos^2r_d\\*
\nonumber \lambda_{5,6}&=&\frac14\sin^2 r_d\cos^2 r_d\left(1\pm\sqrt{1+\frac{4\cos^2 r_d}{\sin^4 r}}\right)\\*
 \lambda_{7,8}&=&\frac14\sin^4 r_d\left(1\pm\sqrt{1+\frac{4\cos^2 r_d}{\sin^4 r_d}}\right)
\end{eqnarray}

As we can see, $\lambda_8$ is non-positive and $\lambda_6$ is negative for all values of $a$, therefore the state will always preserve some degree of distillable entanglement. The negativity is, after some basic algebra
\begin{equation}
\mathcal{N}_d^{AR}=\frac12\cos^2r_d
\end{equation}

\subsubsection{Bipartition Alice-AntiRob}

The eigenvalues of the partial transpose density matrix for the bipartition Alice-AntiRob \eqref{etaAaRd} turn out to be
\begin{eqnarray}\label{eigetaAaRf}
\nonumber \lambda_1&=&\frac12\sin^4 r_d;\qquad \lambda_2=\frac12\sin^2 r_d \cos^2 r_d\\*
\nonumber \lambda_3&=&\frac12\cos^2 r_d;\qquad \lambda_4=\frac12\sin^2 r_d\\*
\nonumber \lambda_{5,6}&=&\frac14\sin^2 r_d\cos^2 r_d\left(1\pm\sqrt{1+\frac{4\tan^2 r_d}{\cos^2 r_d}}\right)\\*
\lambda_{7,8}&=&\frac14\cos^4 r_d\left(1\pm\sqrt{1+\frac{4\tan^2 r_d}{\cos^2 r_d}}\right)
\end{eqnarray}
The negativity, after some basic algebra, turns out to be
\begin{equation}
\mathcal{N}^{A\bar R}_d=\frac12\sin^2 r_d\end{equation}

It is remarkable --and constitutes one of the most suggestive results of this article-- that we have obtained here a conservation law for the entanglement Alice-Rob and Alice-AntiRob, since the sum of both negativities is independent of the accelerations
\begin{equation}\label{conservationN}
\mathcal{N}^{A\bar R}_d +\mathcal{N}_d^{AR} =\frac12
\end{equation}
This is similar to the result \eqref{conservation1} for mutual information. Again, one could check that for spinless fermion fields the same  conservation law \eqref{conservationN} obtained here applies. This was again expected due to the universality principle demonstrated in \cite{Edu3}.

As we will see below, this conservation of quantum correlations is exclusive of fermionic fields. Statistics can be blamed for this neat result. This is in line with what was suggested in \cite{Edu3}, since nothing of the sort will be found for bosonic fields. 

\subsubsection{Bipartition Rob-AntiRob}

The eigenvalues of the partial transpose density matrix for the bipartition Rob-AntiRob \eqref{etaRaRd} turn out to be
\begin{eqnarray}\label{eigetaRaRf}
\nonumber\lambda_{1}&=&\frac{\cos^4 r_d}{2};\qquad \lambda_{2}=\frac{\sin^4 r_d}{2}\\
\nonumber\lambda_{3}&=&\lambda_4=\frac{\sin^2 r_d\cos^2 r_d}{2}\\
\nonumber\lambda_{5,6}&=&\pm\frac{\sin r_d\,\cos^3r_d}{2}\\
\nonumber\lambda_{7,8}&=&\pm\frac{\cos r_d\,\sin^3r_d}{2}\\
\nonumber\lambda_{9,10}&=&\frac{\cos^2 r_d}{4}\left(1\pm\sqrt{1+\sin^2(2r_d)}\right)\\
\nonumber\lambda_{11,12}&=&\frac{\sin^2 r_d}{4}\left(1\pm\sqrt{1+\sin^2(2r_d)}\right)\\
\lambda_{13,14,15,16}&=&\pm\frac{\sin (2r_d)}{8}\left(1\pm\sqrt{1+\sin^2(2r_d)}\right)
\end{eqnarray}
and therefore, the sum of the negative eigenvalues gives a negativity
\begin{equation}\label{negafinalfer}
\mathcal{N}^{R\bar R}_d=\frac{1}{4}\left[\frac{\sin(2r_d)}{2}-1+[1+\sin(2r_d)]\sqrt{1+\sin^2(2r_d)}\right]
\end{equation}

As it can be seen from \eqref{negafinalfer} and graphically in fig. \ref{negaferm}, the entanglement between Rob and AntiRob, created as Rob accelerates, grows up to a finite value. Although this entanglement is useless for quantum information tasks because of the impossibility of classical communication between both sides of an event horizon, the result obtained here is a useful hint in order to understand how information behaves in the proximity of horizons.

Comparing again this result with spinless fermions \cite{AlsingSchul}, we see that for Dirac fields, the maximum value of the negativity is greater. Again this is strongly related with the dimension of the Hilbert space, as we will comment more deeply below, when we deal with scalar fields.

\begin{figure}[h]
\includegraphics[width=.45\textwidth]{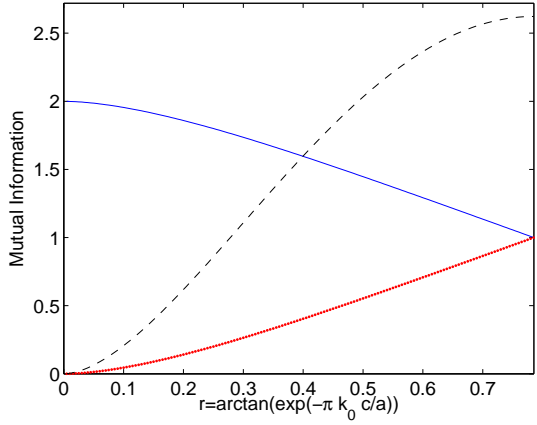}
\caption{(Colour online) Dirac field: Mutual information tradeoff and conservation law between the systems Alice-Rob and Alice-AntiRob as acceleration varies. It is also shown the behaviour of the mutual information for the system Rob-AntiRob. Blue continuous line: Mutual information $AR$, red dotted line: Mutual information $A\bar R$, black dashed line: Mutual information $R\bar R$ }
\label{mutuferm}
\end{figure}

\begin{figure}[h]
\includegraphics[width=.45\textwidth]{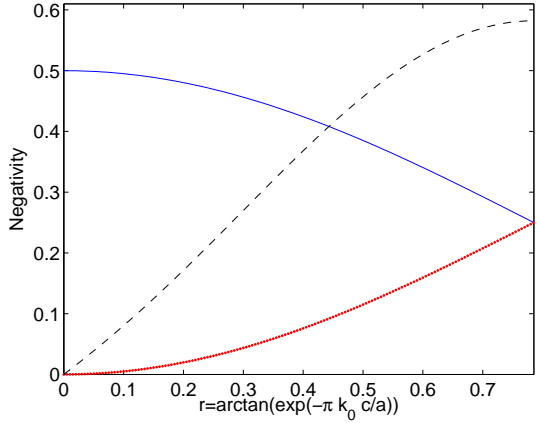}
\caption{(Colour online) Dirac field: Negativity tradeoff and conservation law between the systems Alice-Rob and Alice-AntiRob  as acceleration varies. It is also shown the behaviour of the quantum correlations for the system Rob-AntiRob. Blue continuous line: Negativity $AR$, red dotted line: Negativity $A\bar R$, black dashed line: Negativity $R\bar R$ }
\label{negaferm}
\end{figure}

\section{Correlations for the scalar field}\label{sec5}

The density matrix for the whole tripartite state, which includes modes in both sides of the horizon along with Minkowskian modes, is built from \eqref{entangledsca}
\begin{equation}\label{tripasca}
\rho^{AR\bar R}_s=\proj{\Psi_s}{\Psi_s}
\end{equation}

As in the fermion case, the three different bipartitions for the scalar field case are obtained as follows
\begin{eqnarray}
\label{AR1}\rho^{AR}_s&=&\tr_{IV}\rho^{AR\bar R}_s\\*
\label{AAR1}\rho^{A\bar R}_s&=&\tr_{I}\rho^{AR\bar R}_s\\*
\label{RAR1}\rho^{R\bar R}_s&=&\tr_{M}\rho^{AR\bar R}_s
\end{eqnarray}
and the density matrix for each individual subsystem
 \begin{eqnarray}
\label{A1}\rho^{A}_s&=&\tr_{I}\rho^{AR}_s=\tr_{IV}\rho^{A\bar R}_s\\*
\label{R1}\rho^{R}_s&=&\tr_{IV}\rho^{R\bar R}_s=\tr_{M}\rho^{AR}_s\\*
\label{aR1}\rho^{\bar R}_s&=&\tr_{I}\rho^{R\bar R}_s=\tr_{M}\rho^{A \bar R}_s
\end{eqnarray}

The bipartite systems are characterized by the following density matrices
\begin{eqnarray}\label{rhoars1}
\rho^{AR}_s&=&\sum_{n=0}^{\infty}\frac{\tanh^{2n}r_s}{2\cosh^2 r_s}\Big[\proj{0n}{0n}+\frac{\sqrt{n+1}}{\cosh r_s}\Big(\proj{0n}{1\, n+1}\nonumber\\*
&&+\proj{1\, n+1}{0n}\Big)+\frac{n+1}{\cosh^2 r_s}\proj{1\,n+1}{1\,n+1}\Big]\nonumber\\*
\end{eqnarray}
\begin{eqnarray}\label{rhoa-rs1}
\nonumber\rho^{A\bar R}_s&\!\!\!=\!\!\!&\sum_{n=0}^{\infty}\frac{\tanh^{2n}r_s}{2\cosh^2 r_s}\Big[\!\proj{0n}{0n}\!+\!\frac{\sqrt{n+1}}{\cosh r_s}\tanh r_s\Big(\!\ket{0\,n+1}\\*
&\!\!\!\!\!\!&\times\bra{1 n}+\proj{1 n}{0\,n+1}\Big)+\frac{n+1}{\cosh^2 r_s}\proj{1n}{1n}\Big]
\end{eqnarray}
\begin{eqnarray}\label{rhor-rs1}
\nonumber\rho^{R\bar R}_s&=&\sum_{\substack{n=0\\m=0}}^{\infty}\frac{\tanh^{n+m}r_s}{2\cosh^2 r_s}\Big(\proj{nn}{mm}+\frac{\sqrt{n+1}\sqrt{m+1}}{\cosh^2 r_s}\\*
&&\times\proj{n+1\,n}{m+1\,m}\!\Big)
\end{eqnarray}
where the bases are respectively
\begin{eqnarray}\label{barbolbasis}
 \ket{nm}&=&\ket{n^A}_{M}\ket{m^R}_{I}\\*
\ket{nm}&=&\ket{n^A}_{M}|m^{\bar R}\rangle_{IV}\\*
\ket{nm}&=&\ket{n^R}_{I}|m^{\bar R}\rangle_{IV}
\end{eqnarray}
for \eqref{rhoars1}, \eqref{rhoa-rs1} and \eqref{rhor-rs1}.

On the other hand, the density matrices for the individual subsystems \eqref{A1}, \eqref{R1},\eqref{aR1} are

\begin{equation}\label{Robpartial}
\rho^{R}_s=\sum_{n=0}^\infty\frac{\tanh^{2(n-1)}r_s}{2\cosh^2 r_s}\left[\tanh^2 r_s+\frac{n}{\cosh^2r_s}\right]\proj{n}{n}
\end{equation}
\begin{equation}\label{ARobpartial}
\rho^{\bar R}_s=\sum_{n=0}^\infty\frac{\tanh^{2n}r_s}{2\cosh^2 r_s}\left[1+\frac{n+1}{\cosh^2r_s}\right]\proj{n}{n}
\end{equation}
\begin{equation}\label{AlicedeAliceRob}
\rho^{A}_s=\frac12\left(\proj{0}{0}+\proj{1}{1}\right)
\end{equation}

\subsection{Mutual Information: creation, exchange and conservation}

Mutual information accounts for correlations (both quantum and classical) between two different parts of the system. Its definition is \eqref{mutualdef}.

To compute the mutual information  for each bipartition we will need the eigenvalues of the corresponding density matrices. We shall go through all the process detailedly in the lines below.

\subsubsection{Bipartition Alice-Rob}

The density matrix for the system Alice-Rob \eqref{rhoars1}  consists on an infinite number of $2\times2$ blocks in the basis $\{\ket{0 n},\ket{1\, n+1}\}_{n=0}^\infty$ which have the form
\begin{equation}
\frac{\tanh^{2n}r_s}{2\cosh^2 r_s}
\left(\!\begin{array}{cc}
1 & \dfrac{\sqrt{n+1}}{\cosh r_s}\\
\dfrac{\sqrt{n+1}}{\cosh r_s} & \dfrac{n+1}{\cosh^2r_s}
\end{array}\!\right)
\end{equation}
whose eigenvalues are
\begin{eqnarray}\label{eigAR}
\nonumber\lambda^1_n&=&0\\*
\lambda^2_n&=&\frac{\tanh^{2n}r_s}{2\cosh^2 r_s}\left(1+\frac{n+1}{\cosh^2 r_s}\right)
\end{eqnarray}

\subsubsection{Bipartition Alice-AntiRob}

Excepting the diagonal element corresponding to $\proj{00}{00}$ (which forms a $1\times1$ block itself) the density matrix for the system Alice-AntiRob \eqref{rhoa-rs1} consists on an infinite number of $2\times2$ blocks in the basis $\{\ket{0 n},\ket{1\, n-1}\}_{n=1}^\infty$ which have the form
\begin{equation}
\frac{\tanh^{2n}r_s}{2\cosh^2 r_s}
\left(\!\begin{array}{cc}
1 & \dfrac{\sqrt{n}}{\sinh r_s} \\
\dfrac{\sqrt{n}}{\sinh r_s} & \dfrac{n}{\sinh^2 r_s}\\
\end{array}\!\right)
\end{equation}
We can gather all the eigenvalues in the following expression
\begin{eqnarray}\label{eigAaR}
\nonumber\lambda^1_n&=&\frac{\tanh^{2n}r_s}{2\cosh^2 r_s}\left(1+\frac{n}{\sinh^2r_s}\right)\\*
\nonumber \lambda^2_n&=&0\\*
\end{eqnarray}

\subsubsection{Bipartition Rob-AntiRob}

It is easy to see that the density matrix for Rob-AntiRob  \eqref{rhor-rs1} --which basically consists in the direct sum of two blocks of infinite dimension-- only has range $\operatorname{rank}(\rho^{R\bar R}_s)=2$. Therefore its eigenvalues are zero except for two of them, which are
\begin{eqnarray}\label{eigRaR}
\nonumber\lambda^{R\bar R}_1&=&\sum_{n=0}^{\infty}\frac{\tanh^{2n}r_s}{2\cosh^2r_s}=\frac12\label{lambda1RaR}\\*\label{lambda2RaR}
\lambda^{R\bar R}_2&=&\sum_{n=0}^{\infty}\frac{(n+1)\tanh^{2n}r_s}{2\cosh^4r_s}=\frac12
\end{eqnarray}
So, the Von Neumann entropy for $\rho^{R \bar R}$ is
\begin{equation}\label{entrop}
S^{R\bar R}=1
\end{equation}

\subsubsection{Von Neumann entropies for each subsystem and mutual information}

To compute the Von Neumann entropies we need the eigenvalues of every bipartition and the individual density matrices. The eigenvalues of $\rho^{AR}_s$, $\rho^{A\bar R}_s$, $\rho^{R\bar R}_s$ are respectively \eqref{eigAR}, \eqref{eigAaR} and \eqref{eigRaR}.

The eigevalues of the individual systems density matrices can be directly read from \eqref{Robpartial}, \eqref{ARobpartial} and \eqref{AlicedeAliceRob} since $\rho^R_s$, $\rho^{\bar R}_s$ and $\rho^A_s$ have diagonal forms in the Fock basis. The Von Neumann entropy for a partition $B$ of the system is \eqref{Vonneu}.

At this point, computing the entropies is quite straightforward. Von Neumann entropy for the Rob partial system is
\begin{eqnarray}\label{entropyref}
\nonumber S_R&=&-\sum_{n=0}^\infty\frac{\tanh^{2(n-1)}r_s}{2\cosh^2 r_s}\Big(\tanh^2 r_s+  \frac{n}{\cosh^2 r_s}\Big)\\*
&&\!\!\!\!\!\!\!\times\log_2\!\left[\frac{\tanh^{2(n-1)}r_s}{2\cosh^2 r_s}\Big(\tanh^2 r_s + \frac{n}{\cosh^2 r_s}\Big)\!\right]
\end{eqnarray}
The partial matrices have a similar mathematical structure. Therefore, we can express the non-trivial entropies for the all the possible partitions as a function of the entropy \eqref{entropyref} for Rob's partial system
\begin{eqnarray}
\nonumber S_{\bar R}&=&\frac{S_R}{\tanh^2 r_s}-\frac{1}{2\sinh^2 r_s}\log_2\left(\frac{1}{2\cosh^2 r_s}\right)\\*
\nonumber &&+\log_2\Big(\tanh^2 r_s\Big)\\*
 S_{AR}&=&S_{\bar R}; \qquad \!\! S_{A\bar R}=S_{R}; \qquad \!\! S_{R\bar R}=S_{A}=1
\end{eqnarray} 
Notice that the expression for $S_{\bar R}$ may appear to blow up as $r_s\rightarrow0$, however this is not the case and it can be checked analytically using \eqref{entropyref} that $\lim_{r\rightarrow 0} S_{\bar R} = 0$.

 The mutual information for all the possible bipartitions of the system will be
\begin{eqnarray}
\nonumber I_{AR}&=&S_A+S_R-S_{A R}=1+S_R -S_{\bar R}\\*
\nonumber I_{A\bar R}&=&S_A+S_R-S_{A\bar R}=1+ S_{\bar R}-S_R\\*
\nonumber I_{R\bar R}&=& S_R+S_{\bar R}-S_{R\bar R}=S_R+S_{\bar R}-1
\end{eqnarray}  

Again we obtain a conservation law of the mutual information for the system Alice-Rob and Alice-AntiRob
\begin{equation}\label{conservationbos}
 I_{AR} + I_{A\bar R}=2
\end{equation}
which again suggests a correlations transfer from the system Alice-Rob to Alice-AntiRob as the acceleration increases.

\begin{figure}[h]
\includegraphics[width=.45\textwidth]{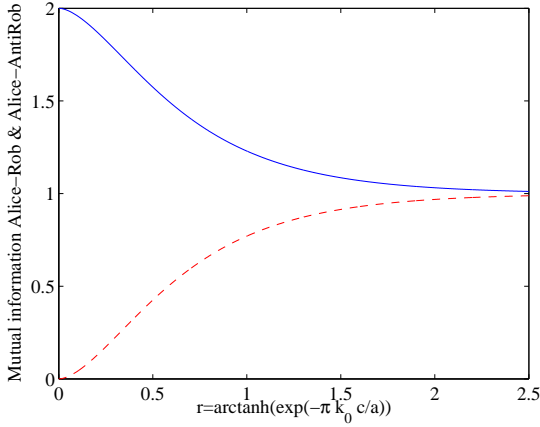}
\caption{(Colour online) Scalar field: Mutual information tradeoff and conservation law between the systems Alice-Rob and Alice-AntiRob. Blue continuous line: Mutual information $AR$, red dashed line: Mutual information $A\bar R$, red dashed line: Mutual information $A\bar R$ }
\label{mututradeoffbos}
\end{figure}

Although the conservation law is the same as for fermion fields \eqref{conservation1}, the specific dependance of the mutual information with the acceleration is different, as it can be seen in fig. \ref{mututradeoffbos}. Later, when we analyse the negativity for all the bipartitions, we will see that, even though mutual information fulfills this conservation law, we must wait for the analysis of quantum correlations to appreciate the striking differences between fermions and bosons.  

\begin{figure}[h]
\includegraphics[width=.45\textwidth]{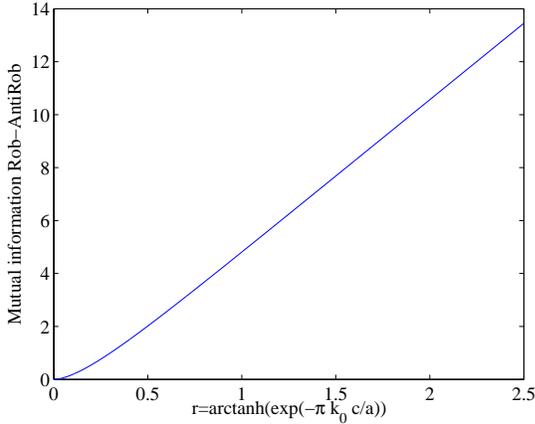}
\caption{(Colour online) Scalar field: Behaviour of the mutual information for the system Rob-AntiRob as acceleration varies}
\label{mutuRARbos}
\end{figure}

Fig.\ref{mutuRARbos} shows how the correlations across the horizon (Rob and AntiRob) increase with no bound as Rob accelerates. This unbounded growing is not only strongly related with the infinite dimension of the Hilbert space, but also very much influenced by statistics. As we will see later, we require  the infinite dimension of the bosons Hilbert space in order to have correlations $R\bar R$ which survive the limit $a\rightarrow\infty$. That is not the case for fermions, where those correlations survive the limit even though the Hilbert space has finite dimension. We will discuss how the infinite dimension is responsible for the unbounded growing of correlations across the horizon after studying the negativity in subsection \ref{negatsec}.

\subsection{Entanglement behaviour across the horizon}\label{negatsec}

 As we did above, we will compute the negativity \eqref{negativitydef} for the scalar case.  To compute it, we need the partial transpose of the bipartite density matrices \eqref{rhoars1}, \eqref{rhoa-rs1} and \eqref{rhor-rs1}, which we will notate as $\eta^{AR}_s$, $\eta^{A\bar R}_s$ and $\eta^{R \bar R}_s$  respectively.

\begin{eqnarray}\label{etaARs}
\eta^{AR}_s&=&\sum_{n=0}^{\infty}\frac{\tanh^{2n}r_s}{2\cosh^2 r_s}\Big[\proj{0n}{0n}+\frac{\sqrt{n+1}}{\cosh r_s}\Big(\proj{0\, n+1}{1n}\nonumber\\*
&&\!\!\!\!\!\!\!+\proj{1 n}{0\,n+1}\Big)+\frac{n+1}{\cosh^2 r_s}\proj{1\,n+1}{1\,n+1}\Big]
\end{eqnarray}
\begin{eqnarray}\label{etaAaRs}
\nonumber\eta^{A\bar R}_s&\!\!\!=\!\!\!&\sum_{n=0}^{\infty}\frac{\tanh^{2n}r_s}{2\cosh^2 r_s}\Big[\proj{0n}{0n}+\frac{\sqrt{n+1}}{\cosh r_s}\tanh r_s\Big(\ket{0 n}\\*
&&\!\!\!\!\!\!\times\bra{1\, n+1}\!+\!\proj{1\, n+1}{0n}\!\Big)\!+\!\frac{n+1}{\cosh^2 r_s}\proj{1n}{1n}\!\Big]
\end{eqnarray}
\begin{eqnarray}\label{etaRaRs}
\nonumber\eta^{R\bar R}_s&=&\sum_{\substack{n=0\\m=0}}^{\infty}\frac{\tanh^{n+m}r_s}{2\cosh^2 r_s}\Big(\proj{nm}{mn}+\frac{\sqrt{n+1}\sqrt{m+1}}{\cosh^2 r_s}\\*
&&\times\proj{n+1\,m}{m+1\,n}\!\Big)
\end{eqnarray}

In the following subsections we shall compute the negativity of each bipartition of the system.
 
\subsubsection{Bipartition Alice-Rob}

Excepting the diagonal element corresponding to $\proj{00}{00}$ (which forms a $1\times1$ block itself), the partial transpose of the density matrix $\rho^{A R}_s $ \eqref{etaARs} has a $2\times2$ block structure in the basis $\{ \ket{0\, n+1},\ket{1 n}\}$
\begin{equation}\label{blocks}
\frac{\tanh^{2n}r_s}{2\cosh^2 r_s}
\left(\!\begin{array}{cc}
\tanh^2 r_s & \dfrac{\sqrt{n+1}}{\cosh r_s}\\
\dfrac{\sqrt{n+1}}{\cosh r_s} & \dfrac{n}{\sinh^2r_s}
\end{array}\!\right)
\end{equation}
Hence, the eigenvalues of \eqref{etaARs} are
\begin{eqnarray}
\nonumber\lambda^1&=&\frac{1}{2\cosh^2r_s}\\*
\nonumber\lambda^2_n&=&\frac{\tanh^{2n} r_s}{4\cosh^2 r_s}\left[\left(\frac{n}{\sinh^2r_s}+\tanh^2 r_s\right)\right.\\*
&&\left.\pm\sqrt{\left(\frac{n}{\sinh^2r_s}+\tanh^2 r_s\right)^2+\frac{4}{\cosh^2 r_s}}\right]
\end{eqnarray}
And then the negativity for this bipartition is
\begin{eqnarray}
\nonumber\mathcal{N}^{AR}_s&=&\sum_{n=0}^\infty\frac{\tanh^{2n} r_s}{4\cosh^2 r_s}\left|\left(\frac{n}{\sinh^2r_s}+\tanh^2 r_s\right)\right.\\*
&&\!\!\!\!\!\!\!\left.-\sqrt{\left(\frac{n}{\sinh^2r_s}+\tanh^2 r_s\right)^2+\frac{4}{\cosh^2 r_s}}\right|
\end{eqnarray}

\begin{figure}[h]
\includegraphics[width=.45\textwidth]{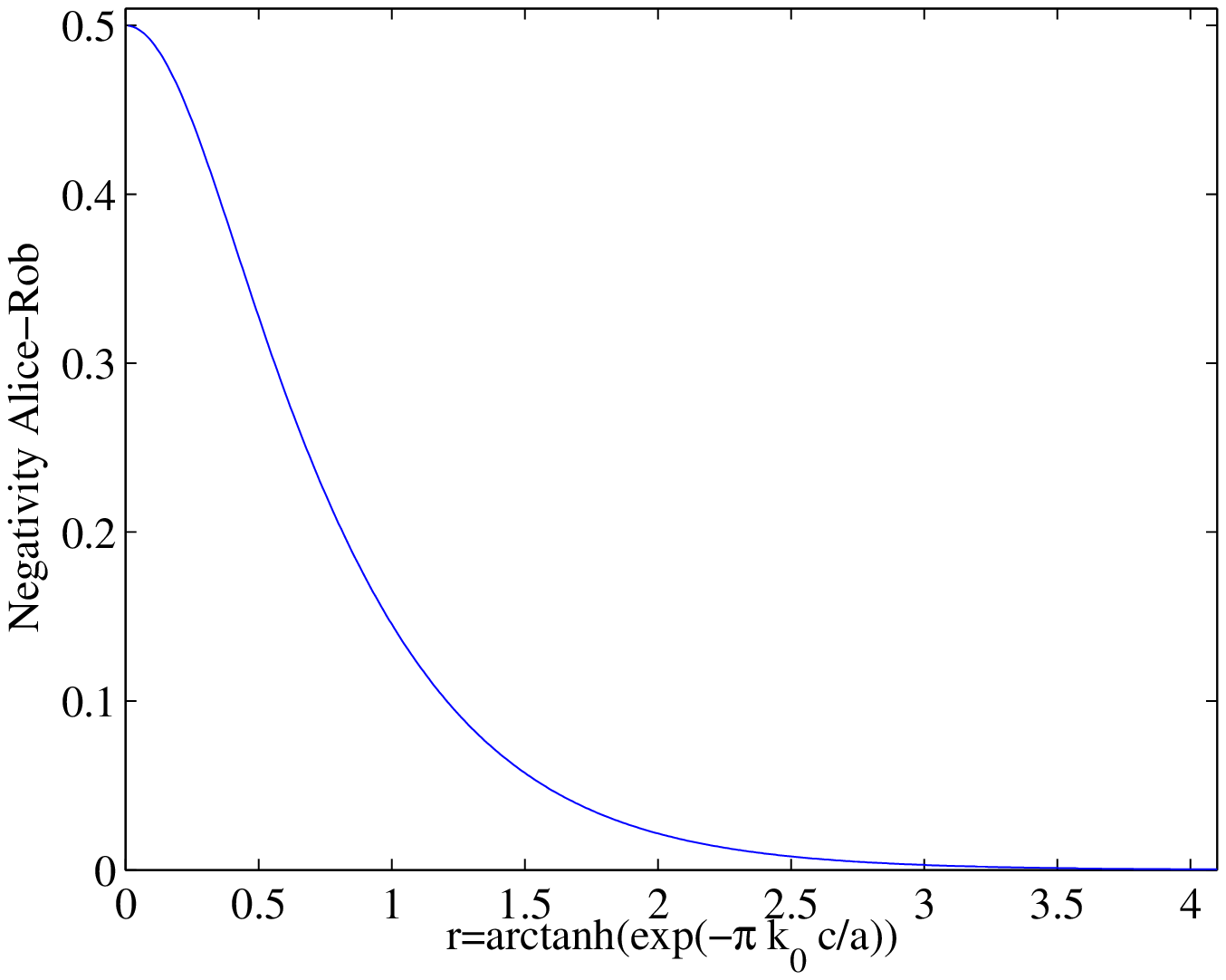}
\caption{(Colour online) Scalar field: Behaviour of the negativity for the bipartition Alice-Rob as Rob accelerates}
\label{negARbos}
\end{figure}

Fig. \ref{negARbos} shows $\mathcal{N}^{AR}_s$ as a function of $r_s$.

\subsubsection{Bipartition Alice-AntiRob}

Excepting the diagonal element corresponding to $\proj{10}{10}$ (which forms a $1\times1$ block itself), the partial transpose of the density matrix $\rho_s^{A \bar R}$ \eqref{etaAaRs} has a $2\times2$ block structure in the basis $\{ \ket{0 n},\ket{1\, n+1}\}$ 
\begin{equation}\label{blocksbosAaR}
\frac{\tanh^{2n}r_s}{2\cosh^2 r_s}
\left(\!\begin{array}{cc}
1 & \dfrac{\tanh r_s}{\cosh r_s}\sqrt{n+1}\\
\dfrac{\tanh r_s}{\cosh r_s}\sqrt{n+1} & \dfrac{\tanh^2 r_s}{\cosh^2r_s}(n+2)
\end{array}\!\right)
\end{equation}
Hence, the eigenvalues of \eqref{etaAaRs} are
\begin{eqnarray}
\nonumber\lambda^1&=&\frac{1}{2\cosh^4 r_s}\\*
\nonumber\lambda^2_n&=&\frac{\tanh^{2n}r_s}{4\cosh^2 r_s}\left[\left(1+(n+2)\frac{\tanh^2 r_s}{\cosh^2 r_s}\right)\right.\\*
&&\!\!\!\!\left.\pm\sqrt{\left(1+(n+2)\frac{\tanh^2 r_s}{\cosh^2 r_s}\right)^2-\frac{4\tanh^2 r_s}{\cosh^2 r_s}}\right]
\end{eqnarray}
Therefore, the negtivity for this bipartition is always $0$, independently of the value of Rob's acceleration. The important implications of this striking result are discussed below.

\subsubsection{Bipartition Rob-AntiRob}

The partial transpose of the density matrix $\rho^{R\bar R}_s$  \eqref{etaRaRs} has a block structure, but the blocks themselves are of different dimensions  which grow up to infinity. Because of this, negativity is not as easily computable as for the other cases, not being possible to write it in a closed form.

However it is still possible to compute the eigenvalues of \eqref{etaRaRs} numerically taking into account that the blocks which form the matrix are  endomorphisms which act in the subspace expanded by the basis $B_{D}=\{\ket{mn}\}$ in which $m+n=D-1=\text{constant}$, which is to say, the fisrt block acts within the subspace expanded by the basis $B_1=\{\ket{00}\}$, the second $B_2=\{\ket{01},\ket{10}\}$, the third $B_3=\{\ket{02},\ket{20},\ket{11}\}$, the fourth $B_4=\{\ket{03},\ket{30},\ket{12},\ket{21}\}$ and so forth. In this fashion, the whole matrix is an endomorphism within the subspace $\bigoplus_{i=1}^\infty S_i$ being $S_i$ the subspace (of dimension $D=i$) expanded by the basis $B_i$.

Let us denote $M_D$ the blocks which form the matrix \eqref{etaRaRs}, being $D$ the dimension of each block. Then its structure is

\begin{equation}\label{blockss}
M_D=\left(\!
\begin{array}{cccccccc}
0  & a_1  & 0 & 0 & \cdots & \cdots& \cdots& 0 \\
a_1 & 0 & a_2 & 0 & \cdots & \cdots& \cdots & 0\\
0 & a_2 & 0 & a_3 & \cdots & \cdots& \cdots& 0\\
0 & 0 & a_3 &0 & a_4 & \cdots& \cdots& 0\\
0 & 0 & 0 &  \ddots &\ddots &  \ddots &\cdots& 0\\
\vdots  & \vdots  & \vdots  & \vdots  & \ddots  & \ddots  & \ddots  & \vdots \\
0 & 0 & 0 &  0 &\cdots&  \ddots &0& a_{D-1}\\
0 & 0 & 0 &  0 &0&  \dots &a_{D-1}& a_{D}\\
\end{array}\!\right)
\end{equation}
which is to say, the diagonal terms are zero except for the last one, and the rest of the matrix elements are zero excepting the two diagonals on top and underneath the principal diagonal. The elements $a_n$ are defined as follows
\begin{equation}
a_{2l+1}=\frac{(\tanh r_s)^{D-1}}{2\cosh^2 r_s}
\end{equation}
\begin{equation}
a_{2l}=\sqrt{D-l}\,\sqrt{l}\frac{(\tanh r_s)^{D-2}}{2\cosh^4r_s}
\end{equation}
Notice that the elements are completely different when the value of the label $n$ is odd or even.

\begin{figure}[h]
\includegraphics[width=.45\textwidth]{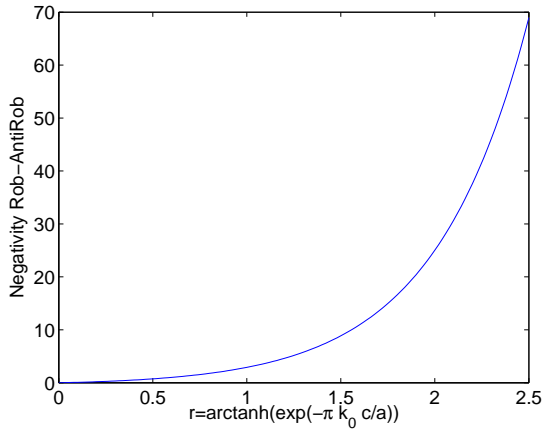}
\caption{(Colour online) Scalar field: Behaviour of the negativity for the bipartition Rob-AntiRob as Rob accelerates}
\label{negaRARbos}
\end{figure}

As the whole matrix is the direct sum of the blocks
\begin{equation}
\eta_s^{R \bar R}=\bigoplus_{D=1}^\infty M_D
\end{equation}
the eigenvalues and, specifically, the negative eigenvalues of $\eta_s^{R \bar R}$ would be the negative eigenvalues of all the blocks $M_D$ gathered togheter. It can be shown that the absolute value of the negative eigenvalues of the blocks decreases quickly as the dimension increases. Thus, the negativity $\mathcal{N}^{R\bar R}_s$ converges promtply to a finite value for a given value of $r_s$.  Fig \ref{negaRARbos} shows the behaviour of $\mathcal{N}^{R\bar R}_s$ with $r_s$, showing that the entanglement increases unboundedly between Rob and AntiRob.

Let us compare these results with the fermion case. 

First, as it was shown in \cite{AlsingSchul}, the negativity of the system Alice-Rob decreases as Rob accelerates, vanishing in the limit $a\rightarrow\infty$, instead of remaining finite as in the fermionic cases \cite{AlsingSchul,Edu2}.

What may be more surprising is the behaviour of quantum correlations of the system Alice-AntiRob. In the fermion case negativity grows monotonically from zero (for $a=0$) to a finite value (for $a=\infty$). Nevertheless, for scalars, Alice-AntiRob negativity is identically zero for all acceleration. Hence, there is no transfer of entanglement from Alice-Rob to Alice-AntiRob as it was the case for fermions. Still, correlations (classical) are not lost as it can be concluded from \eqref{conservationbos}.

Why do we obtain such loss of entanglement for the bosonic case and, conversely, it does not happen in the fermionic case? The answer is, once again, statistics. 

One could think of the infinite dimensionality of the Hilbert space for scalars (compared to the finite dimension for fermions) as the cause of this different behaviour. However we shall prove that it has to do with the bosonic nature of the field rather than with the infinite dimensionality of the Hilbert space. We can see this considering hardcore bosons instead of scalars, which is to say, we can limit the occupation number for the bosonic modes to a certain finite limit $N$ instead of taking $N\rightarrow\infty$. By doing so, we transform the infinite dimension Hilbert space for bosons into a finite dimension one. As we will see below, in any case, the negativity would continue being zero for all $a$.

To illustrate this argument, we could compare the fermionic case with the hardcore bosonic case with $N=2$ which would be its analogous, namely, a bosonic field whose occupation number is limited to $2$. Looking at the partially transposed density matrix for the fermionic case, we can seek for the negative eigenvalues origin and, then, compare the analogous structure that would appear for the hardcore bosonic case.  We will see that, in the latter case, no negative eigenvalues are obtained.

Recalling the form of the matrix \eqref{etaAaRd}, we can see that it has a structure of four $1\times1$ blocks, which give positive eigenvalues, and two $2\times2$ blocks, which are the ones whose eigenvalues contribute to the negativity. These two blocks are 
\begin{equation}
\frac12\left(\!\begin{array}{cc}
\cos^4 r_d & \sin r_d\cos^2 r_d\\[1.8mm]
\sin r_d\cos^2 r_d & 0\\
\end{array}\!\right)
\end{equation}
in the basis $\left\{\ket{00},\ket{\uparrow\uparrow}\right\}$ and
\begin{equation}
 \frac12\left(\!\begin{array}{cc}
\sin^2 r_d \cos^2 r_d & -\sin^3 r_d\\[1.8mm]
-\sin^3 r_d& 0\\
\end{array}\!\right)
\end{equation}
in the basis $\left\{\ket{0\downarrow},\ket{\uparrow\pa}\right\}$.

For hardcore bosons with occupation number limited to $n=N$ the structure would be quite similar to \eqref{etaAaRs}, but this time we would have two $1\times1$ blocks instead of one (the elements $\proj{00}{00}$ and $\proj{NN}{NN}$) and $N$ $2\times2$ blocks with the structure \eqref{blocksbosAaR} but with $n=0,\dots,N-1$. Specifically, for $N=2$ the two blocks are
\begin{equation}\label{blocksbosAaR2}
\frac{1}{2\cosh^2 r_s}
\left(\!\begin{array}{cc}
1 & \dfrac{\tanh r_s}{\cosh r_s}\\[2.5mm]
\dfrac{\tanh r_s}{\cosh r_s} & \dfrac{2\tanh^2 r_s}{\cosh^2r_s}
\end{array}\!\right)
\end{equation}
in the basis $\left\{\ket{00},\ket{11}\right\}$ and
\begin{equation}\label{blocksbosAaR3}
\frac{\tanh^{2}r_s}{2\cosh^2 r_s}
\left(\!\begin{array}{cc}
1 & \dfrac{\sqrt{2}\tanh r_s}{\cosh r_s}\\[2.5mm]
\dfrac{\sqrt{2}\tanh r_s}{\cosh r_s} & \dfrac{3\tanh^2 r_s}{\cosh^2r_s}
\end{array}\!\right)
\end{equation}
in the basis $\left\{\ket{01},\ket{12}\right\}$.

The key difference comes from the second diagonal term of the blocks, which in the fermionic case it is impossible to obtain due to the peculiar structure that fermionic statistics imposes on \eqref{onepart2}. On the other hand, for bosons this term is non-zero, and furthermore it has a value large enough to prevent the partial transposed density matrix  from having negative eigenvalues.

As mentioned in section \ref{sec3}, this result is independent of the election of the particular value for the spin components for Alice and Rob in \eqref{entangleddir}. It also does not depend on $N$ (the limit for occupation number we impose to hardcore bosons modes) and this phenomenon is exclusively ruled by statistics. Besides, this is an evidence of the difference between hardcore bosons and fermions, which sometimes in the literature are considered as the same. This is not the case, at least speaking about quantum correlations, about which we have observed that statistics is a key feature.

This is additional confirmation of the important role that statistics plays in the behaviour of correlations in non-inertial frames, and specifically in the proximity of black holes. In fact, getting closer and closer to the Rindler horizon (i.e. $a\rightarrow\infty$) the situation resembles more and more the scenario in which Rob is arbitrarily close to the event horizon of a Schwartzschild black hole, while Alice is free-falling into it.

\begin{figure}[h]
\includegraphics[width=.45\textwidth]{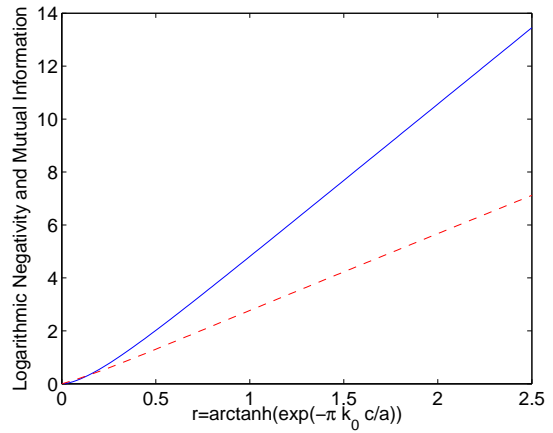}
\caption{(Colour online) Scalar field: Comparison of growth of quantum and all (quantum+classical) correlations for the system Rob-AntiRob as acceleration increases. Quantum correlations are accounted for by logarithmic negativity. This figure compares this entanglement measurement with mutual information. Blue continuous line represents mutual information and red dashed line represents Logarithmic negativity}
\label{logneg}
\end{figure}

About the bipartition Rob-AntiRob, at first glance at fig \ref{mutuRARbos} and \ref{negaRARbos} one could think that there might be some inconsistency between the behaviour of entanglement and mutual information, as the latter grows linearly while negativity seems to grow exponentially. Since mutual information accounts for all the correlations (quantum and classical) between Rob and AntiRob, the result may appear paradoxical. However this apparently inconsistent results are due to the fact that negativity cannot be identified as the entanglement itself, but as a monotone which grows as the degree of entanglement does.  The specific functional form chosen for the monotone is not imposed by physical motivations. Actually, we could have chosen logarithmic negativity  --instead of negativity-- as our entanglement monotone since it is in fact better to be compared with mutual information due to its additivity properties \cite{logneg}. The result obtained in this case, shown in fig. \ref{logneg}, is that when acceleration grows both growths become linear.

Coming back to the argumental line of this paper, we observe that entanglement grows unboundedly for this bipartition, conversely to the fermion case in which negativity increases up to a certain finite limit as Rob accelerates. This different behaviour is related not only with the dimension of the Hilbert space but also it is strongly influenced by statistics. 

Actually, if we consider again hardcore bosons we can see from \eqref{blockss} that limiting the dimension would give a finite number of blocks which contribute to the negativity, being the last one truncated and having dimension $N-1$ instead of $N+1$. In any case, taking the limit $a\rightarrow\infty$ we can see that negative eigenvalues of the partially transpose density matrix tend to zero as $a$ grows. Therefore negativity vanishes in that limit.

Naturally, the same happened in the standard bosonic case, but then we had that negativity was the sum of an infinite number of terms each vanishing when $a\rightarrow\infty$. The negativity resulted divergent, although. Now we are adding only a finite number of vanishing terms so that negativity, which was divergent  when $a\rightarrow\infty$ for standard bosons, now vanishes in such limit. This behaviour contrasts with the fermionic case and the standard scalar field case.

This and other points related with the impact of dimension and statistics on this correlations across the horizon will appear elsewhere.

\section{Conclusions}\label{conclusions}

This work focused on the bipartite correlations between different spatial-temporal domains in the presence of an event horizon. Specifically, we analyse all the possible bipartitions of an entangled system composed by an inertial observer and an accelerated one, who inhabits a universe with an event horizon.

As pointed out in the introduction, Rindler space-time is the simplest case which, presenting a horizon, also reproduces two interesting physical scenarios. Namely, it describes the point of view of an uniformly accelerated observer and, in the limit $a\rightarrow\infty$, it reproduces an observer resisting arbitrarily close to a Swchartzschild  black hole event horizon.

First of all, we have studied the relation between the behaviour of entanglement of the system Alice-Rob and Alice-AntiRob, which are the two bipartitions to which one could assign physical meaning. We have shown that the dimension of the Hilbert space has little to do with the entanglement behaviour.

We recall two results to support this statement, on one hand, in \cite{Edu3} we studied the case of a Dirac and a spinless fermion field beyond the single mode approximation in which fermionic Hilbert space turned out to have infinite dimension. On the other hand, here we have analysed bosons in a Hilbert space of infinite and finite dimensions (scalar field and hardcore scalar respectively). The results are that the characteristic differences between fermion fields and bosonic fields for the bipartitions $AR$ and $A\bar R$ manifest independently of the dimension of the Hilbert space.

Here we have disclosed a great difference in the behaviour of quantum correlations for fermions and bosons. In the fermionic case we showed that, at the same time as Unruh decoherence destroys the entanglement of the system Alice-Rob, entanglement is created between Alice and AntiRob. This means that the quantum entanglement lost between Alice and the field modes outside the event horizon is gained between Alice and the modes inside. This is expressed through the entanglement conservation law \eqref{conservationN}, which we have deduced for fermions.

Nevertheless, for bosonic states it was shown that, as acceleration increases, entanglement is quickly and completely lost between Alice and Rob while no quantum correlations are created between Alice and AntiRob. Moreover, no entanglement of any kind survives among any physical bipartition of the system in the limit $a\rightarrow\infty$ for the bosonic case. This contrasts with the fermionic case where the amount of entanglement among all the physical bipartitions of the system remains always constant.

This goes in the line of the previous work \cite{Edu3} in which we suggested that all the entanglement which survives a black hole event horizon is purely statistical, and gives no further information than the statistical correlations which inescapably appear due to the fermionic nature of the field. 

These powerful and suggesting results may well be related with what was obtained in previous works as \cite{sta1,sta2}. In those works it was demonstrated, for pure states in first quantization, that the fact of considering indistinguishable fermions implies that there are quantum correlations of which we cannot get rid due to the antisymmetrisation of the fermionic wavefunctions. Of course, the scenario here is not the same, but analysing together the results obtained here and those in \cite{sta1,sta2}  reveals the somewhat special feature of fermionic fields regarding entanglement comportment: statistics information, or in this case, information of the fermionic nature of the field, cannot be erased from our system, and still, that information is expressed by means of quantum correlations.

Another remarkable result is the conservation law for mutual information for fermions \eqref{conservation1} and bosons \eqref{conservationbos} shown in  in fig. \ref{mutuferm} and fig. \ref{mututradeoffbos}. The detailed  behaviour is different for both cases but the same conservation law is obtained for the mutual information of the bipartitions Alice-Rob, and Alice AntiRob. Mutual information accounts for both classical and quantum correlations (despite the fact that in general there is no trivial relation between negativity and mutual information). However for the bosonic case, the mutual information distributes between the systems Alice-Rob and Alice-AntiRob more rapidly than for the fermion case.

This result for mutual information means that correlations are always conserved for the systems Alice-Rob and Alice-AntiRob, despite the fact that quantum correlations vanish for the bosonic case and they are preserved (and this is the effect of statistics) in the fermionic case. The fact that classical correlations behave in a similar way for fermions and bosons while quantum correlations comport so differently suggests again that the quantum entanglement which survives the black hole limit is merely statistical.

Another difference between fermion and bosons appears analysing the correlations between modes inside and outside the horizon. It is interesting to notice that, as the non-inertial partner accelerates, correlations across the horizon are created. Both, dimension of the Hilbert space and statistics, play a fundamental role in the behaviour of these correlations. We have obtained that for Dirac fields these correlations, quantum and classical, grow as Rob accelerates up to a finite value at the limit $a\rightarrow\infty$. This limit is greater than the analogous limit  obtained for spinless fermions in \cite{AlsingSchul} whose Hilbert space for each mode is smaller. For the bosonic case, on the contrary, those correlations grow unboundedly, being infinite when $a\rightarrow\infty$. Surprisingly, for the hardcore bosonic field, in which Hilbert space have an arbitrarily large but finite dimension, the behaviour of this correlations is not even monotonical, being zero for the inertial limit as well as for the infinite acceleration regime, but non-zero in between. Therefore, statistics and dimensionality must be taken into account in order to find the origin of this comportment.

We have dealt with this bipartition separately since it cannot be given a physical interpretation in terms of information theory. However, it is worthwhile to analyse its comportment since it could reveal the relative roles of dimensionality and statistics when accounting for the behaviour of correlations in the presence of an event horizon. This topic, which deserves further study, is expected to appear elsewhere.

It is important to recall that the limit $a\rightarrow\infty$ can be understood as considering an observer moving in a trajectory arbitrarily close to the event horizon of a Schwarzschild black hole \cite{Alicefalls}. So, along with the interest of describing the correlations between accelerated observers, this study gives further insight  about the fate of correlations in the presence of a black hole. Together with \cite{Edu3}, our results could be of use to tackle the problem of the information paradox in black holes, for which they indicate that statistics plays a  very important role.

\section{Acknowledgments}

This work was partially supported by the Spanish MICINN Project FIS2008-05705/FIS and by the CAM research consortium QUITEMAD S2009/ESP-1594. E. M-M was partially supported by a CSIC JAE-PREDOC2007 Grant.

\bibliographystyle{apsrev}

\end{document}